\documentclass[sigconf]{acmart}

\AtBeginDocument{%
  \providecommand\BibTeX{{%
    \normalfont B\kern-0.5em{\scshape i\kern-0.25em b}\kern-0.8em\TeX}}}

\copyrightyear{2021}
\acmYear{2021}
\setcopyright{acmcopyright}\acmConference[CHI '21]{CHI Conference on Human Factors in Computing Systems}{May 8--13, 2021}{Yokohama, Japan}
\acmBooktitle{CHI Conference on Human Factors in Computing Systems (CHI '21), May 8--13, 2021, Yokohama, Japan}
\acmPrice{15.00}
\acmDOI{10.1145/3411764.3445497}
\acmISBN{978-1-4503-8096-6/21/05}

\usepackage{graphicx}
\usepackage[utf8]{inputenc}
\usepackage{amsmath}
\usepackage{units}
\usepackage{xcolor}
\usepackage{booktabs}
\usepackage{textcomp}
\usepackage{multirow}

\usepackage{microtype}        % Improved Tracking and Kerning
\usepackage{ccicons}          % Cite your images correctly!

\def\plaintitle{Adapting User Interfaces with Model-based Reinforcement Learning}
\def\plainauthor{Kashyap Todi, Gilles Bailly, Luis A. Leiva, Antti Oulasvirta}
\def\plainkeywords{Adaptive User Interfaces; Reinforcement Learning; Predictive Models; Monte Carlo Tree Search}

\author{Kashyap Todi}
\email{kashyap.todi@gmail.com}
\affiliation{%
  \institution{Aalto University}
  \country{Finland}
 }
\author{Gilles Bailly}
\email{gilles.bailly@sorbonne-universite.fr}
\affiliation{%
  \institution{Sorbonne Universit\'e, CNRS, ISIR}
  \country{France}
}
\author{Luis A. Leiva}
\email{luis.leiva@uni.lu}
\affiliation{%
  \institution{University of Luxembourg}
  \country{Luxembourg}
 }
\author{Antti Oulasvirta}
\email{antti.oulasvirta@aalto.fi}
\affiliation{%
  \institution{Aalto University}
  \country{Finland}
 }

\DeclareMathOperator{\EX}{\mathbb{E}}% expected value

\newcommand{\compresslist}
{
    \setlength{\itemsep}{1pt}
    \setlength{\parskip}{0pt}
    \setlength{\parsep}{0pt}
}

\definecolor{linkColor}{RGB}{6,125,233}
\hypersetup{%
  pdftitle={\plaintitle},
  pdfauthor={\plainauthor},
  pdfkeywords={\plainkeywords},
  pdfdisplaydoctitle=true, % For Accessibility
  bookmarksnumbered,
  pdfstartview={FitH},
  colorlinks,
  citecolor=black,
  filecolor=black,
  linkcolor=black,
  urlcolor=linkColor,
  breaklinks=true,
  hypertexnames=false
}

\begin{document}

\title[Adapting User Interfaces with Model-based Reinforcement Learning]{Adapting User Interfaces with Model-based \\Reinforcement Learning}

\begin{abstract}
Adapting an interface requires taking into account both the positive and negative effects that changes may have on the user.
A carelessly picked adaptation may impose high costs to the user -- for example, due to surprise or relearning effort -- or ``trap'' the process to a suboptimal design immaturely.
However, effects on users are hard to predict as they depend on factors that are latent and evolve over the course of interaction.
We propose a novel approach for adaptive user interfaces that yields a conservative adaptation policy:
It finds beneficial changes when there are such and avoids changes when there are none.
Our model-based reinforcement learning method plans sequences of adaptations and consults predictive HCI models to estimate their effects.
We present empirical and simulation results from the case of adaptive menus, showing that the method outperforms both a non-adaptive and a frequency-based policy.
\end{abstract}

\begin{teaserfigure}
  \centering
  \includegraphics[width=\textwidth]{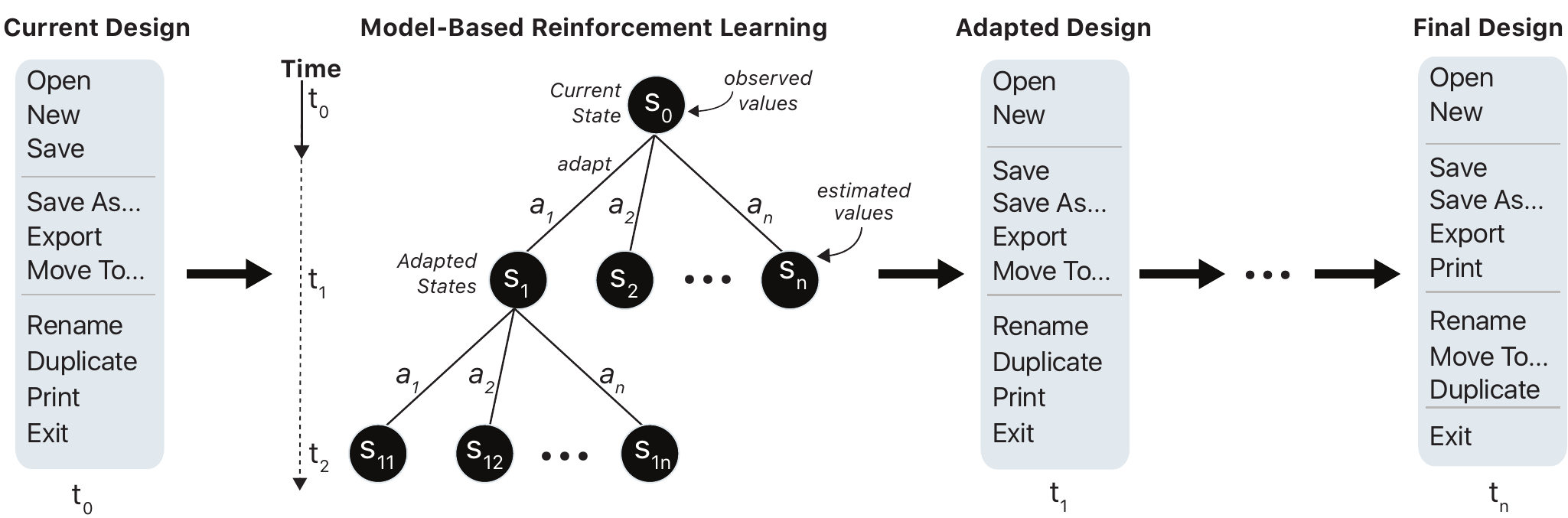}
  \caption{
    We present a model-based reinforcement learning approach for adaptive UIs that can improve usability while avoiding unexpected changes that surprise the user or require relearning.
    An interface is adapted by simulating several possible \emph{sequences} of adaptations and evaluating them using predictive models in HCI.
    This approach avoids greedy, disadvantageous adaptations, and may anticipate possible user responses even with limited observation data.
  }
   \Description{A figure with five parts, connected using left-to-right arrows. The first part shows a linear menu as the `current design'; the second part shows a tree diagram illustrating model-based reinforcement learning; the third part shows an adapted version of the menu; the fourth part has ellipsis to indicate truncation; and the fifth part shows another adapted menu as the 'final design'.}
  \label{fig:adapt_plan}
\end{teaserfigure}




\maketitle

\section{Introduction}

Adaptive user interfaces can autonomously change the content, layout, or style of an interface to improve their fit with the user's capabilities and interests.
This paper looks at a foundational technical problem of adaptive interfaces that lies at the intersection of human--computer interaction and machine learning research:
\emph{How to select adaptations?}
An adaptive system must decide \emph{what} to adapt, and \emph{when} -- or \emph{when not} -- to make changes.
Different computational approaches to this problem have been studied, among others, rule-based systems, heuristics, bandits, Bayesian optimisation, and supervised learning (see \autoref{sec:relwork}).
Although positive empirical results have been obtained (e.g., ~\cite{sears_split_1994, tsandilas_bubbling_2007, cockburn_predictive_2007, findlater_ephemeral_2009, bailly_visual_2017, todi_familiarisation:_2018, 10.1145/3241381, 10.1145/3025171.3025207}),
known approaches have been criticised for being unpredictable and unreliable; they pick detrimental adaptations unacceptably often \cite{lavie_benefits_2010,10.1145/985692.985704,greenberg_adaptive_1985,leiva_restyling_2012,Vertegaal03_aui}.

\emph{Estimation of utility} is required for selecting an adaptation.
Utility -- in this case -- refers to the usefulness of an adaptation to the user, or how it is perceived to benefit interaction when possible costs are taken into account.
Picking an adaptation can be considered a hypothesis on how useful it is for user.
Unfortunately, utility is very hard to estimate accurately -- hard both at design time as well as interactively from the kind of data these systems have access to, such as clicks or viewing duration.
In machine learning terms, utility is \emph{latent}.
Moreover, in adaptive interaction, utility is also \emph{non-stationary}.
That is, the skills and interests of the user evolve over time.
A change that would make sense in the beginning when the user is novice with the design may be devastating for an experienced user.

We believe that adaptive systems could provide greater benefits by \emph{planning sequences of adaptations} that gracefully lead a user through gradual changes.
However, non-stationarity makes planning challenging:
considering only a short period of time (i.e. a short horizon) can result in suboptimal designs.
An adaptation that is overfit to a novice could be impossible to recover from later on when the user is more experienced.
On the other hand, planning a long sequence of adaptations increases the size of search space, growing exponentially with the length of the planning horizon.

\emph{Model-based reinforcement learning} is here developed as a principled and effective approach to these issues.
We define the adaptation problem as a \emph{stochastic sequential decision problem} \cite{bellman1970decision}, where the adaptive system should plan a sequence of adaptations over a long horizon.
Reinforcement learning (RL) is a class of machine learning methods appropriate for this type of problems.
Typically, a policy is learnt via trial-and-error that maximises future cumulative utility.
Our RL approach is \emph{model-based} as it uses predictive HCI models to estimate utility.
These models simulate consequences -- benefits and costs -- of possible adaptation sequences without actually executing them \cite{kaelbling1996reinforcement,sutton2018reinforcement}.
However, their application is conditioned on their accuracy: They should accurately predict short-term costs such as due to relearning as well as longer-term improvements to performance.
Generally, such models are available at least in the areas of pointing, menu interaction, and graphical layouts.
In comparison with an alternative, the \emph{model-free} approach, the model-based approach requires less training and better generalises across conditions \cite{kaelbling1996reinforcement,sutton2018reinforcement}.
However, finding the best adaption -- by assessing the value of each sequence of adaptations with predictive models -- is computationally costly, especially when considering sequences of changes over a long horizon.
To solve this computational problem in an online setting, we use a combination of Monte Carlo Tree Search (MCTS) for planning, and deep learning to boost performance.
To avoid extensive trial-and-error with users in the loop, our deep learning models are trained offline using HCI models.

Our general approach can be applied for various HCI applications, such as adaptive mobile homescreens, graphical layouts, and application menus.
We demonstrate it specifically in the context of \emph{adaptive menus}.
The task is to adapt the interface by changing the arrangement and grouping of menu items (Figure 1), thus improving the menu's usability.
We exploit and extend multiple menu search models from literature to estimate the upper and lower bounds of the value of an adaptation as well as their change as the user learns.
This helps us form an adaptation policy that accounts for different user strategies and avoids adaptations that incur high costs to users.
Our technical evaluation shows that our solution can tackle realistic problem sizes, and find favourable adaptations, on a commodity computer.
Finally, we present results from an empirical evaluation where the approach compared favourably against a non-adaptive baseline design and a frequency-based adaptation policy.

To sum up, this paper makes three key contributions:
\begin{enumerate}\compresslist
    \item A new formulation for adaptive interfaces that formalises them as a stochastic sequential decision-making problem;
    \item Development of model-based reinforcement learning for planning adaptations in the case where users learn;
    \item Application in adaptive menus with demonstrated benefits to usability.
\end{enumerate}

\section{Previous work: Machine Learning Methods for Adaptive Interfaces}\label{sec:relwork}

Our work contributes to methods for adaptive interfaces designed to operate autonomously,
that is without explicit feedback or training samples from user.
The core computational problem we review here is how to pick an adaptation;
we do not cover issues like prior elicitation, explainability, nor the design space of intelligent interaction techniques.

\subsection{Rules, heuristics, and logic}
Early work on this topic studied rule systems, heuristics, and logic as the basis of deciding what to adapt (see, e.g., \cite{puerta1994model}).
For example, in menu-based interaction, most systems still follow a heuristic approach, where adaptation is picked based on hand-written heuristics that exploit click frequency, visit duration, recency or other specific features that can be computed from observation data \cite{gobert_sam:_2019}.
These approaches, in general, are feasible only when sensed input is highly predictive of the most appropriate adaptation.
Another limitation is that writing a comprehensive and accurate rule system requires plenty of foresight.
A rule system must be developed that, on the one hand, covers conceivable conditions the system can enter and, on the other, can graciously resolve conflicts when multiple rules apply.

\subsection{Machine learning}

The prevailing understanding is that \emph{learning} is a key capability for adaptive systems \cite{langley1997machine}.
Two learning capabilities are needed:
\emph{(1) inference}, the capability to update assumptions about the user based on observations;
and \emph{(2) decision-making}, the capability to choose appropriate adaptation in the light of assumptions about the user \cite{oulasvirta2018computational}.
The two challenges can be relaxed, for example if user state is trivially known,
or if the state is highly predictive of appropriate adaptation.
In the latter case, the problem can be approached as a \emph{supervised learning} problem, where a mapping is learned between user data and suitable adaptations.
While this approach has been successful for input techniques, such as gesture recognition \cite{10.1007/3-540-46616-9_10},
it is an open question if this scales up to adaptive interfaces,
which must not only learn user state but flexibly decide how to intervene in the user interface.
A practical obstacle is how to obtain a dataset that describes the consequences of possible adaptations on possible users.
% To circumvent this issue, mixed-initiative techniques have been recently used to recommend a set of adaptations using a machine learning model \cite{doi:10.1080/10447318.2020.1824742}.

In the rest of the section, we focus on the general case, where both inference and decision-making are required and non-trivial.

\subsection{Bandit systems and Bayesian optimisation}

\emph{Bandit systems} are one of the most successful probabilistic approaches to this problem, not only for recommendation systems but also for interface design and adaptation ~\cite{lomas2016interface}.
Each adaptation is modelled as an `arm' associated with a distribution describing expected gains.
Given prior data and new evidence on the measured success of an adaptation, bandits use Bayes theorem to update expectation.
Importantly, a principled solution is offered to the exploration/exploitation problem.
Methods like Thompson sampling can optimally balance between exploring actions, to learn about which actions work, and exploitation, to converge to good designs.

\emph{Bayesian optimisation} generalises bandit systems to the case of multiple interrelated decision variables \cite{shahriari2015taking}.
It is a global optimisation method that tries to find optimal adaptation by probing to a black box function; here, the user.
It is a robust and sample-efficient and well-suited for noisy, expensive-to-evaluate functions.
The method uses a \emph{surrogate model} for approximating the model fit across the parameter space and quantify uncertainty.
This is necessary for the acquisition rule to address the exploration--exploitation trade-off.
This way, it is possible to learn adaptive responses via trial and error.
Applications have been shown in human-in-the-loop design of interface features \cite{dudley2019crowdsourcing} and adaptation of low-dimensional design features \cite{kim2020optimal}.
However, while bandits and BO have been successful in simpler adaptation problems, like recommendations and calibration of interface parameters, their intrinsic limitation is that they are myopic; that is, they do not plan over a series of changes -- a capability we need in adaptive interaction.

\subsection{Reinforcement learning}

Unlike bandits, reinforcement learning permits learning policies for \emph{sequences} of actions where rewards are not immediately achievable.
While applications have been shown for example in crowdsourcing \cite{dai2013pomdp}, dialogue systems \cite{young2013pomdp}, and gaze-based interactions \cite{10.1145/3332165.3347933}, a known limitation with the prevailing model-free RL approach is still the extensive amount of poor attempts (or trials) that are required to learn a good policy \cite{sutton2018reinforcement}
This makes it poorly suitable for situations with very large state-action spaces.

\emph{Model-based} RL uses a predictive model to simulate possibilities without first trying them out, which is useful for adaptive interfaces, because it significantly improves the efficiency of finding good solutions \cite{kaelbling1996reinforcement}.
A policy can be determined with much fewer trials, and if the model is good, at times directly.
Outside of user interfaces, we find applications of model-based RL, for example in board games, robots, video games \cite{kaiser2019model}, as well as behaviour-change applications, such as in generating behavioural instructions for people with dementia \cite{hoey2007assisting}.
A grand obstacle for applications in adaptive systems is the model: where to get a good one?
A related problem is that of \emph{drift}: prediction errors can have a compounding, cumulative effect on planning performance.
In HCI, although previous work has explored the use of model-free RL (e.g., see \cite{leino_computer-supported_2019}),
model-based RL has not been explored in adaptive interfaces at large, as far as we know.
Also, while predictive models have been used for one-shot design generation \cite{oulasvirta2020combinatorial}, design space exploration \cite{10.1145/2901790.2901817},  and for selecting a single action in a myopic manner \cite{jokinen2017modelling, todi_familiarisation:_2018},
they have not been used for simulation-based planning in an adaptive system.

In this paper, we approach the fundamental problem of selecting user interface adaptations by applying model-based RL, and exploiting predictive HCI models, to simulate and plan adaptations.

\section{Problem Formulation}\label{sec:problemdef}

We formulate the problem of adaptation as a \emph{stochastic sequential decision-making problem} \cite{bellman1970decision}.
The adaptive system must \emph{decide} what to adapt, if anything, given its observations.
It should pick a \emph{sequence} of adaptations in order to maximise their expected value to user over a longer window of interactions.
For example, in our application example later on,
we optimise for performance improvements in menu selection tasks achievable over multiple sessions of interactions.
Further, in a \emph{stochastic} problem, the world is neither fully known nor under the control of the system.
In our case, while the system can change the interface, it cannot change the human, which has its own latent processes.
For example, humans learn and change interests.
This complicates the problem:
Any greedily chosen adaptation may lead to irreversibly poor interactions later on.
Thus, adaptations must be picked with a horizon of such developments in mind.

In the following, we formulate this problem as a \emph{Markov decision process} (MDP).
The benefit of an MDP formulation is that it offers a rigorous and actionable understanding of the problem.
In particular, it (1) illuminates the decision problem, (2) links it to a body of theoretical results and practical approaches in AI and ML research, and (3) points toward appropriate algorithmic solutions.

The problem of adaptive interfaces is that of maximising expected cumulative discounted rewards $r(s_t, a_t)$ from acting according to an optimal policy $\pi^*$ (see e.g. \cite{janner2019trust}):
\begin{equation}
\pi^* = \underset{\underset{p(s_{t+1} | s_t, a_t)}{\pi}}{\mbox{argmax}} \EX_\pi \left[\sum\limits_{t=0}^\infty \gamma^{\;t} r(s_t, a_t) \right]
\end{equation}
where,
\begin{itemize}\compresslist
\item \emph{$s \in S$} is a state of interaction;
    This consists of both the interface design ($d$) and the user ($u$)
\item \emph{$a \in A$} is an adaptation; i.e. a change that can be carried out on the interface.
\item \emph{$p$} is a transition function; it provides the probability of transitioning from state $s$ to state $s_{t+1}$ after performing adaptation $a$; i.e., $p(s_{t+1} | s_t, a_t)$.
\item \emph{$r$} is a reward collected for making adaptation $a$ in state $s$.
\item \emph{$\gamma$} is a discount factor controlling for how much to favour immediate (small $\gamma$) vs. long-term (large $\gamma$) reward.
\end{itemize}

Consider the case of adapting the homescreen layout of smartphones, consisting of a grid of application icons.
Here, a state $s$ of the system consists of both the homescreen design $d$ and latent state of the user $u$ who is interacting with the device.
More specifically, the design $d$ can encapsulate factors such as the arrangement of icons, their grouping or relationship to other icons, and other relevant features.
With regard to the user $u$, aspects such as expertise, interests, and abilities, can be considered.

Given an initial homescreen design, with which the user has interacted, an adaptation $a$ would result in a new design by, for example, changing the layout or ordering of icons.
Upon adaptation, the transition function $p$ specifies how the internal state of the user (e.g. their expertise) changes along with the external design state.
The reward $r$ then signifies the benefit an adaptation provides to the user by improving future interactions, for example, by reducing the time required to select an icon.
Finally, the discount factor $\gamma$ indicates to the adaptive system how immediate benefits and long-term improvements contribute towards the reward.
Given this setting, the goal of the system is to find a suitable policy $\tau$ that can be used to select adaptations that maximise the estimated cumulative reward.

Finding a policy to select adaptations can be challenging for adaptive interfaces.
While the true state of the design is fully observable, the system does not have access to the true state of the user.
We can only estimate it through HCI models, where predictions provide us a \emph{belief} about the user status (for the theoretical implications of this, see \cite{araya2010pomdp}).
Further, due to the lack of user feedback, computing reward is not straightforward.
Instead, predictive HCI models can be used to build objective functions related to performance and re-learning costs.
Note that this formulation does not take a stance on what is the `right' objective function.
Finally, at any given state, there is a large number of possible adaptations that can change the design.
When considering a sequence of adaptations over a long horizon, the state space grows exponentially.
In the following, we describe how these challenges are addressed via model-based RL.

\section{Method: Deep Model-based Reinforcement Learning}

The core of this approach considers planning:
the selection of a sequence of adaptation with the goal of maximising utility to user.
Planning algorithms such as minimax and A-star, among others, utilise a tree representation of the search space. They have found several applications (e.g. game-playing \cite{brockington1996taxonomy}, circuit-routing \cite{10.1007/978-3-030-05054-2_14}, etc.).
The tree consists of nodes, connected by branches, representing valid states and transitions between them.
However, classic tree search algorithms often require expansion of the entire tree.
This is computationally expensive given the large number of possible adaptations (breadth) and long horizons (depth) in adaptive interfaces.

\emph{Monte Carlo Tree Search (MCTS)} has been successfully employed in various game-playing applications to plan a sequence of moves efficiently (see \cite{6145622}).
% Unlike minimax, which requires expansion of the entire tree before it can decide moves \gb{not understable at his stage. Reviewer does not know what is minimax and you never mentioned tree before},
MCTS can operate under uncertainty by analysing the most promising moves,
and expanding tree nodes using random sampling based approaches.
A well-known, inspiring application of MCTS is in AlphaGo \cite{silver2016mastering}, the computer program capable of playing the game of Go competitively against human players.
A key insight here has been to incorporate neural networks to help predict which branches have highest expected value,
and thereby deal with larger problem instances.
In our work, we incorporate a value neural network to compute longer sequence cases faster.

\subsection{Planning with MCTS}\label{subsec:mcts}

\begin{figure*}
    \centering
    \includegraphics[width=0.98\textwidth]{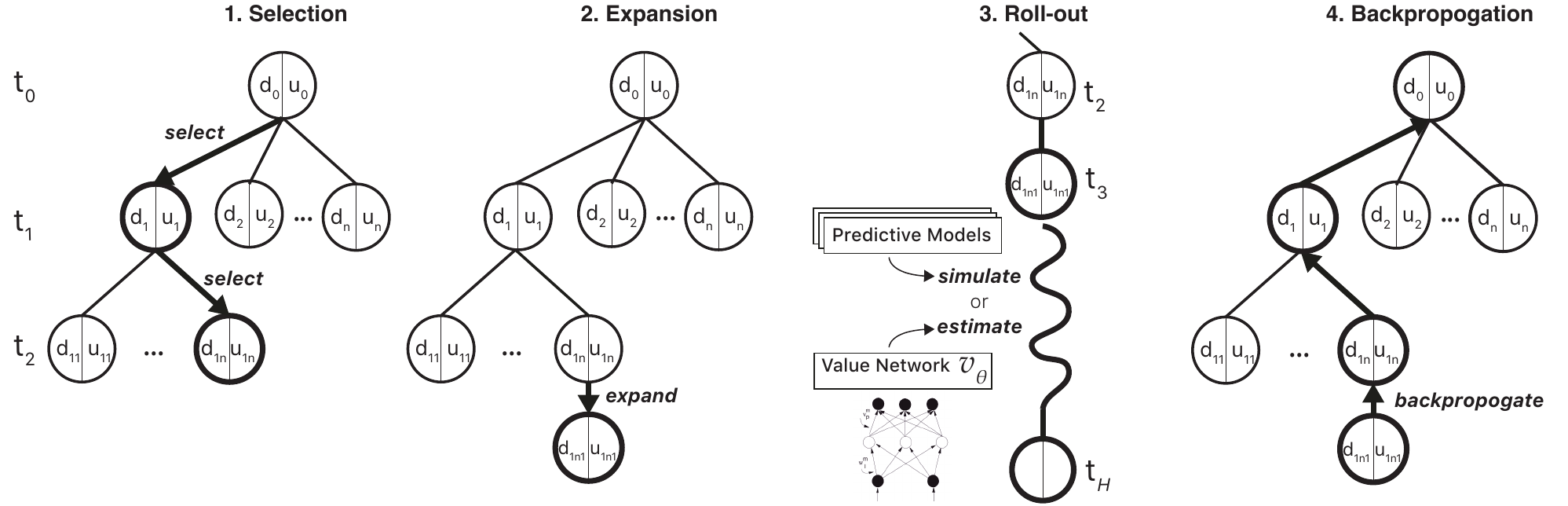}
    \caption{Model-based planning of adaptations using Monte-Carlo Tree Search (MCTS). Adaptations are selected using upper-confidence trees (UCT). After expanding a new node (adaptation), reward estimates are obtained through roll-outs, and backpropogated to the root node.
    The child with the highest value is picked as the next adaptation.}
    \label{fig:mcts}
     \Description{A figure illustrating the four steps of Monte Carlo Tree Search. Each part has a tree diagram highlighting the corresponding phase: selection, expansion, rollout, and backpropogation.}
\end{figure*}

In contrast to game-playing applications, where a win/loss defines the terminal state,
our case does not have a well-defined horizon.
The user could keep interacting for years.
We thus need to estimate a cumulative reward over a horizon of reasonable length.
In other parts, our solution follows standard implementations of MCTS (\autoref{fig:mcts}):

\textbf{1. Selection}: When a user concludes an interaction session,
the root state $s_0$ is given by the current design $d_0$ and user observations $u_0$.
A state $s_j \in S$ is selected via Upper Confidence Trees (UCT),
a widely used estimator in model-based planning \cite{10.1007/11871842_29}.
A key feature is that it has a coefficient $C$ to balance between exploration and exploitation when evaluating several possible UI adaptations:
\begin{equation}
\text{UCT} = \frac{r_j}{n_j} + C \sqrt{\frac{\ln n_i}{n_j}}
\label{eq:uct}
\end{equation}
where $r_j$ is the total reward for child state $s_j$,
$n_j$ is the number of times $s_j$ has been visited,
and $n_i$ is the number of times parent state $s_i$ has been visited.
Exploration constant $C$ in our application is set to $1/\sqrt 2$ following convention.
If all adaptations from the selected state $s_j$ have been previously explored,
then selection is repeated until a leaf state is selected with unexplored adaptations.

\textbf{2. Expansion}: The selected node $s_j$ is next expanded by picking an adaptation $a \in A$ that results in a new state $s_{j+1}$.
At this point, $n_{j+1}, r_{j+1} = 0$.
In our application, we further assume that the expanded state is either visible or invisible to the user.
Invisibility can be exploited to plan multiple adaptations in a single turn.

\textbf{3. Roll-outs and simulations}:
During one \emph{roll-out}, as the tree has no value estimates to inform the selection of consequential states, adaptations $\{a_0, \dots, a_N\} \in A$ are chosen at random, and rewards are estimated using predictive HCI models.
All models simulate what would happen with that adaptation sequence and return an estimate of values over the whole window.
This is repeated for a fixed number of steps, given by the horizon $H$, and cumulative rewards are computed for each predictive model:
\begin{equation}
    r_{j+1} = \sum_{k=j+2}^H r_{k}
\end{equation}{}

\textbf{4. Backpropagation}:
At the end of a simulation, the cumulative reward $r_{j+1}$ is backpropagated from the newly-expanded state $s_{j+1}$ to the initial state $s_0$, and values (rewards $r$ and visits $n$) are updated.

The above steps (1--4) are repeated several times to obtain value estimates for each adapted state.

\textbf{Selecting the next adaptation}:
Given these value estimates, the system can now choose the best adaptation (by setting $C=0$ in \autoref{eq:uct})
to maximise expected utility for the user while avoiding costly changes.

We support several ways for selecting adaptation that allows controlling for the trade-off between risk and gain.
Our approach assumes that there are multiple predictive HCI models that bound the true behaviour, for example by offering best-case and (realistic) worst-case estimates.
Alternatively, predictive models can be included to address multiple objectives, such as task completion time, cognitive load, and disruption \cite{10.1145/1502650.1502691}.
Combining value estimations from all models, we can implement different objective functions, such as:
\begin{enumerate}\compresslist
\item \emph{Average:} The mean of rewards from each model gives total reward $r$, thus accounting for varying user strategies.
\item \emph{Optimistic:} The system assumes optimal user strategy, and selects the model that maximises rewards.
\item \emph{Conservative:} The system ensures that no user strategy is harmed by selecting the lowest-possible reward, and penalising negative rewards.
\end{enumerate}

\subsection{Value Estimation with Deep Neural Networks}

To address large problem sizes in online settings,
where repeating a sufficient number of MCTS simulations to attain robust estimates is infeasible,
we develop a deep neural network architecture that can efficiently provide predictions in real-time.
In place of roll-outs, where simulations can be costly for longer horizons,
a pre-trained \emph{value} network is used to directly obtain value estimates for unexplored states.

\begin{figure}
    \centering
    \includegraphics[width=0.95\linewidth]{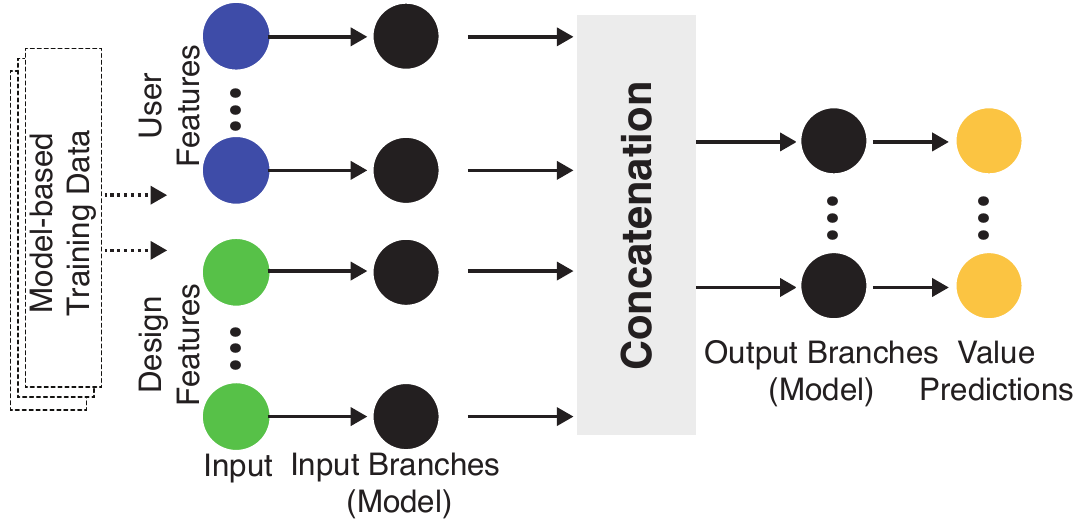}
    \caption{Neural network architecture for obtaining value estimates. Training data is generated using HCI models.}
    \label{fig:value_network_general}
     \Description{An architecture diagram of the neural network for value estimates. The figure shows training data being passed as input, from which user and design features are extracted and passed to input model branches. These branches (represented by circles) are passed to a concatenation box, resulting in output branches and value predictions.}
\end{figure}

As illustrated in \autoref{fig:value_network_general}, we propose an $m$-headed $n$-tailed architecture that is trained end-to-end with backpropagation.
Each of the $m$ input parameters is treated as an independent model branch (head)
that is eventually concatenated and passed to $n$ independent model branches (tails).
During offline training, model-based data is generated using MCTS roll-outs from randomly sampled initial states.
Value estimates, along with state (design and user) information,
are given as input samples to a deep neural network model.
Elementary design and user features are parameterised with the neural network model.
This is then used to predict value estimates for any given state in an online setting.
Note that we decouple value estimations from objective functions (see above),
leaving it up to the adaptive system to decide how to use information from multiple models.
The main advantages of using neural networks for
estimation are that they have high learning capacity,
and can evaluate thousands of states in real-time without running expensive simulations.

\subsection{Applications in Adaptive Interfaces}
\begin{figure*}[t!]
    \centering
    \includegraphics[width=0.97\textwidth]{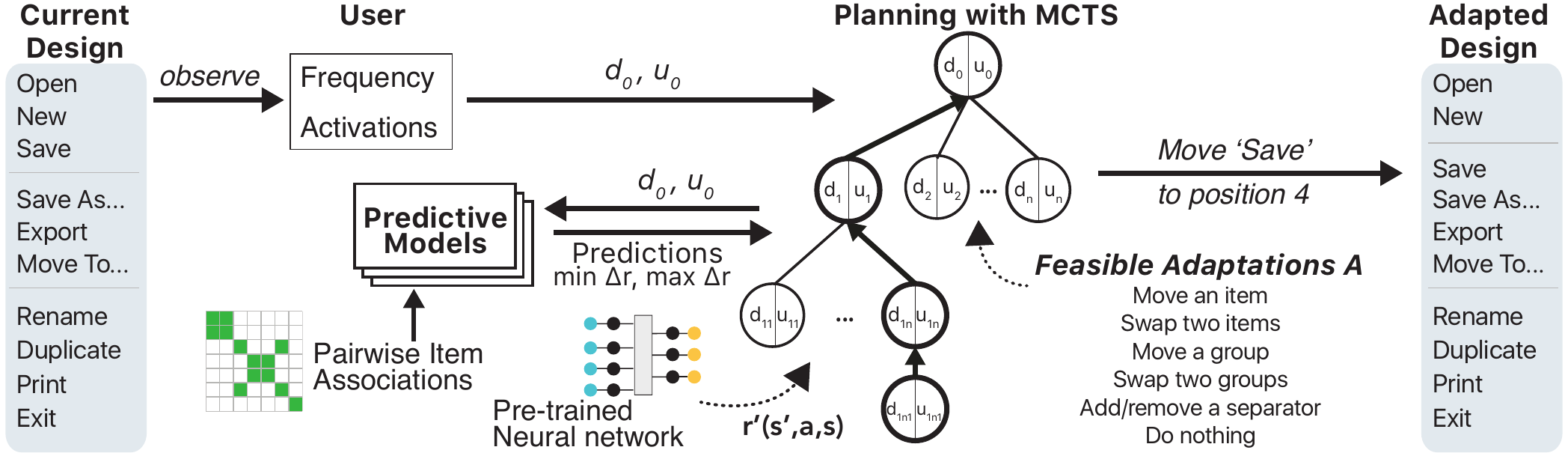}
    \caption{Menu adaptation with deep model-based RL. Given a current design and user observations, MCTS-based planning is used to select the next adaptation. Reward estimates are obtained either using predictive models or from a neural network.}
    \label{fig:menu_mdp}
     \Description{An overview diagram to illustrate the process of menu adaptation using deep model-based RL. The left side shows a current menu design. An arrow from this design points to user observations, which are then passed to the MCTS algorithm. Here, predictive models or the neural network is used to plan adaptations. By selecting feasible adaptations, an adapted menu design is obtained as output.}
\end{figure*}

With certain conditions we outline here, the approach we outline is broadly useful across applications of adaptive interfaces.
For example, it can be used to adapt the structure of webpages layouts, arrange icons on mobile homescreens, or reorganise application menus.
Depending on the application, the goal or objective can differ and might include minimising selection time, increasing saliency, reducing cognitive load, increasing engagement, or a combination of them.
The approach can handle different types of adaptations including the presentation of the graphical elements (position, size, colour, etc.) or their behaviour (number of elements, animations, etc.).

For any application case, a core aspect of our approach relies on HCI models \cite{oulasvirta2020combinatorial} to sufficiently accurately capture the effects of these adaptations on the chosen goals.
For instance, Fitts' law \cite{doi:10.1207/s15327051hci0701} can be used to adapt the location and the size of an elements to minimise pointing time.
Other practical limitations include the size of state--action space.
In the following section, we demonstrate our approach with an application in adaptive menus, and present a novel HCI model to predict selection time in menus.

\section{Application: Adaptive Menus}\label{sec:menuapp}

To demonstrate the applicability of our approach, we tackle a challenging and open question in the field of adaptive interfaces: \emph{adaptive menus}.
Menus have received extensive attention in HCI research because they are widely used, and adaptation has potential to improve usability \cite{vanderdonckt_2019, bailly_visual_2017}.
It is known that unexpected changes in menus can introduce a temporary performance drop, increase cognitive load, and potentially lead to the rejection of the adaption/techniques \cite{vanderdonckt_2019, bailly_visual_2017}.
No general solution has been proposed for autonomous adaptation that could not only move items to the top, but handle reorganisations more comprehensively.
We provide a general solution here considering linear menus with up to 20 items, which covers a wide number of menus typically found in common operating systems, applications, and websites~\cite{bailly_2008_flower,bailly_visual_2017,bailly_menuoptimizer_2013}.
As menu adaptations, we only consider those modifying the position and grouping of items in the menus, leaving other presentation adaptations, such as highlighting and split menus, for future work.
While we consider linear menus with textual labels, our solution can be extended to address the problem of adaptive homescreens (introduced in \autoref{sec:problemdef}) by extending the menu search model to consider two-dimensional grids and graphical icons.

\subsection{Problem Definition}

Following the general formulation in \autoref{sec:problemdef}, we first define the adaptive menu problem.
\autoref{fig:menu_mdp} presents an exemplary illustration.

\textbf{State ($S$):}
A state $s \in S$ gives information about the menu design and the user.

We define a design $d \in D$ as a pair  $<L,M>$ where $L$ is a non-hierarchical linear menu \cite{norman1991psychology} containing an ordered list of items. An item is either a word or a separator (used to create semantic groups).
$M$ is an association matrix defining the \emph{semantic relatedness} between menu items.
In our implementation, $M$ is given by the designer as binary relationships by specifying lists of related items.
For common words, it can be inferred using word embedding models~\cite{pennington2014glove}.

The system observes user clicks on menu items to approximate a user's \emph{expertise} and \emph{interest}.
We use a simplified version of the learning component from ACT-R~\cite{anderson2014atomic} to compute \emph{user expertise} for each menu item $i_l$ in a menu with $n$ items:
\begin{equation}
    B(i_l) = \sum_{j=1}^n ( T - T_{j, i_l} )^{-\rho}
\end{equation}
where $B(i_l)$ is the level of activation of item $i$ at location $l$ in the memory, $T$ is the current time, $T_{j, i_l}$ is the time of the $j^{th}$ selection of $i_l$, and $\rho$ is a decay parameter equal to $0.5$.
\emph{User interest} (or prediction scheme \cite{vanderdonckt_2019}) is given by the frequency distribution of commands selected during the previous interaction session, containing $N$ clicks.
Additional statistical models of user interest and expertise \cite{frensch_composition_1994, fitchett_accessrank_2012, todi_familiarisation:_2018, gobert_sam:_2019}, can be plugged into our architecture.

\textbf{Feasible Adaptations ($A$):}
The set of possible adaptations $A$, through which a menu can be reorganised, includes (1)~moving a menu item to a certain position, (2)~swapping two items, (3)~adding or removing a separator, (4)~moving an entire group, (5)~swapping two groups,  and (6)~not making any changes.

\textbf{Transition function ($p$):}
We use MCTS, where the probability of making an adaptation from state $s_0$ to $s_1$ is given by UCT (\autoref{eq:uct}).
During planning, we balance between exploiting high-reward adaptations and exploring others.
When selecting an adaptation $a$ to make on the current design $d_0$, resulting in a new design $d_1$, a greedy strategy is chosen:
\begin{equation}
    \underset{a}{{\mbox{argmax}}}\  \frac{r_j}{n_j}
\end{equation}

\textbf{Reward ($r$):}
We extend predictive HCI models of menu use to obtain reward estimates.
A key feature of these models is to take into account the implicit cost of adaptations.
Given a pair of menu designs and estimates of user expertise and interest, these models predict selection time for items for varying user strategies.
For each model, the reward $r$ then is the difference in average selection time, weighted by user interest.
Rewards from multiple models can be combined using any of the objective functions given in section \ref{subsec:mcts}.

When an adapted state is assumed to be displayed (visible) to the user, we simulate an interaction session (new clicks) based on user interest, and update expertise accordingly.
Conversely, the system can use invisible states to withhold presentation and combine multiple adaptations, such that the state is used only as a pathway to another state without being displayed. Here, no user updates are applied.

\subsection{Models for Simulating Menu Search}

\begin{figure*}
    \centering
    \includegraphics[width=0.99\textwidth]{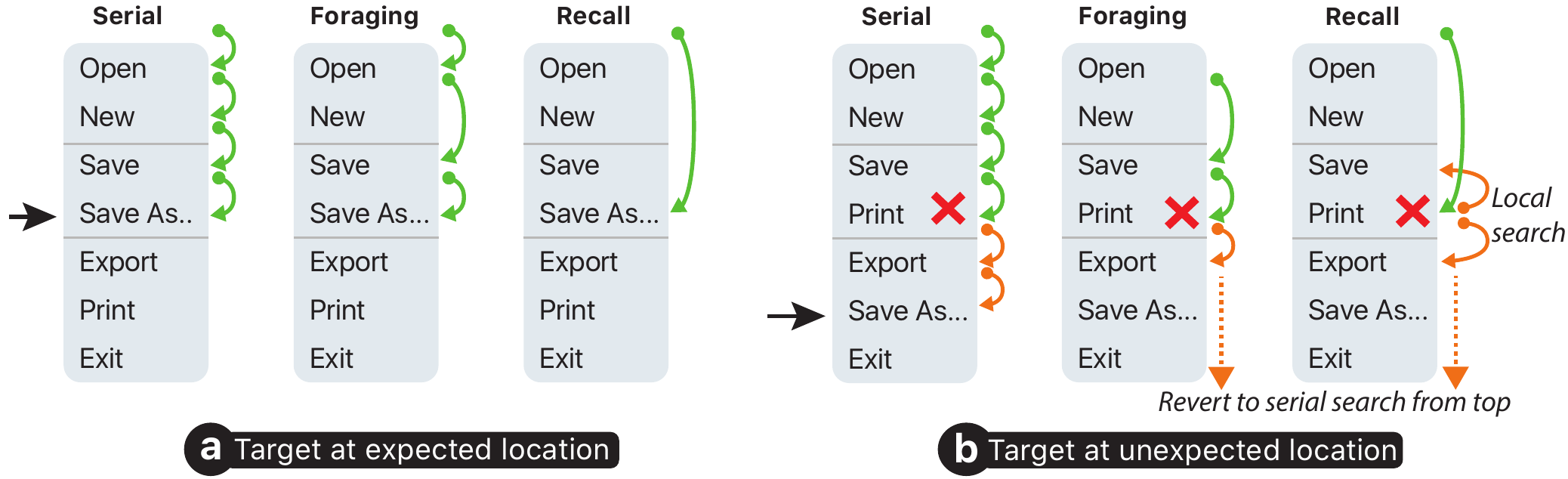}
    \caption{Model-based simulation of search for a target item (\emph{'Save As...'}).
    (a)~when the target is at an expected location, search proceeds as expected;
    (b)~when the target is at an unexpected location, a penalty is imposed upon not finding the target at its expected location (indicated by $\textcolor{red}{\times}$), and search reverts to slower strategies.}
    \label{fig:menumodels}
     \Description{A two-part figure showing model-based simulations of the process of searching for a target menu item. In the first part, the target is at the expected location. For each of the three models (serial, foraging, and recall), arrows indicate how the user searches the menu to find the target. In the second part, the target is at an unexpected location. Green arrows show the normal search behaviour. Upon failure to find the target, orange arrows indicate searching with slower strategies.}
\end{figure*}

Several predictive models explain how users search within linear menus \cite{bailly_14, byrne_2001, cockburn_09, hornof_97, norman1991psychology}.
We build on these models to define three search strategies, and use these to evaluate the utility of a menu design by simulating search tasks, illustrated in \autoref{fig:menumodels}:
\begin{enumerate}\compresslist
    \item \emph{Serial search}: The user searches serially, from top to bottom, until the desired item is found in the menu.
    \item \emph{Foraging search}: Grouping of items is exploited such that the user only searches for the target within relevant groups.
    \item \emph{Recall search}: The user relies on memory to search for the desired item at an expected location in the menu.
\end{enumerate}
Our choice of the three models was made with the hypothesis that they would provide bounds for best-case and worst-case performance.
In the beginning, best-case performance would be governed by foraging and serial search, but as experience accumulates, the (rational) user would shift to foraging based and recall-based strategies.
But making the assumption that a user \emph{might} -- for reasons like lack of effort or interest -- use a poor strategy allows us to define a conservative policy for adaptation that is unlikely to annoy them.
We note that it is possible to plug in additional search strategies (e.g. Random search~\cite{norman1991psychology}) without modifying the general architecture of the algorithm.

\begin{table}[hbt!]
    \caption{Key notations used in this section.}
    \centering
    \begin{tabular}{r|p{0.75\linewidth}}
        \textbf{Notation} & \textbf{Description}\\
        $i_l$ & Target item $i$ at location $l$\\
        $T_\text{model}(i)$ & Search time for an item $i$ with a given model\\
        $\delta$ & Constant for cautious inspection cost of an item\\
        $T_c$ & Constant surprise penalty when an item is not found as expected\\
        $T_\text{trail}$ & Constant pointing time when the cursor trails eye gaze\\
        $M(i_k,i_l)$ & Boolean relationship between items $i_k$ and $i_l$,
        from association matrix $M$\\
        $B(i_l)$ & Activation level for $i$ at $l$
    \end{tabular}
    \label{tab:notation}
\end{table}

\subsubsection{\textbf{Serial search}}
When searching for a menu item $i_e$ at an expected position $e$, serial search \cite{norman1991psychology,cockburn_predictive_2007, bailly_14} consists of a top-to-bottom inspection of the menu, until the item is reached.
Inspection (or reading) cost for any item $i_l$ is given by:
\begin{equation}
    T_\text{read}(i_l) = \frac{\delta}{1 + B(i_l)}
    \label{eq:read}
\end{equation}
where $B(i_l)$ is the activation of $i$ at $l$, and $\delta$ is the cautious inspection cost constant when there is no activation. The total serial search time for a target at expected location $e$ is thus given by:
\begin{equation}
 T_\text{serial}(i_e) =  \sum_{j=1}^{e} T_\text{read}(i_j) + T_\text{trail}
 \label{eq:serial}
\end{equation}
Where $T_\text{trail}$
is a constant pointing time assuming that the mouse cursor trails the eye-gaze (tracking strategy) during serial search \cite{bailly_14,byrne_2001}.

In an adapted menu, if the item's new location $a$ is \textbf{before} the expected location $e$ ($a \leqslant e$), the search time reduces following \autoref{eq:serial}.
However, if $a$ is \textbf{after} $e$ ($a \geqslant e$),
\emph{surprise penalty} $T_c$ is imposed upon not finding the item as expected.
Following this, the user cautiously inspects the remaining items at a slower rate $\delta$.
The total search time is:
\begin{equation}
 T_\text{serial}(i_a) =  T_\text{serial}(i_e) + T_c + (a - e) \cdot \delta + T_\text{pointing}
 \label{eq:serial-slow}
\end{equation}

To support serial search, it is advantageous for an adaptive system to move frequently-used items towards the top.

\subsubsection{\textbf{Foraging search}}
Here, semantic structure (grouping) is exploited to avoid wasting time inspecting groups that most likely do \emph{not} contain the target item \cite{bailly_visual_2017}.
The \emph{anchor}, or first element of a group, ``signals'' what is in the group.
If the anchor is unrelated to the target, the user skips the group.
If the anchor is related, the user performs a serial search within this group.

Consider a menu where $n_g$ is the number of groups, $n_g(i_e)$ is the location of the group that contains the target ($i_e$), and $e$ the expected target location within the group.
$M(g_j, i_e) \in [0,1]$ specifies if $i_e$ is associated to an anchor $g_j$. In a well-organised menu, where the anchor of one of the groups is related to the target item, foraging search time is:
\begin{flalign}
\label{eq:forage}
 T_\text{forage}(i_e)
    & =
    \sum_{j=1}^{n_g(i_e)}
    \left( \vphantom{\int}
        T_\text{read}(g_j) \right. + \\ \nonumber
    &
        \left. M(g_j, i_e) \cdot
        \sum_{k=1}^{\min(e(g_j), s(g_j))} T_\text{read}(i_k)
    \right) + T_\text{trail}
\end{flalign}
% \begin{equation}
% \label{eq:forage}
%  T_\text{forage}(i_e)  =
%     \sum_{j=1}^{n_g(i_e)}
%     \left(
%         T_\text{read}(g_j) +
%          M(g_j, i_e) \cdot
%         \sum_{k=1}^{\min(e(g_j), s(g_j))} T_\text{read}(i_k)
%     \right) + T_\text{trail}
% \end{equation}
where $s(g_j)$ is the size of the group, and $e(g_j)$ is the location of the target item $t_e$ within group $g_j$ if it is located within the group, $\infty$ otherwise.
Finally, $T_\text{trail}$ remains the constant pointing time when the mouse cursor trails the eye gaze.
Thus, items within related groups are inspected serially until the target is successfully found.

In a poorly organised menu, where the target item is not located within the expected group(s), a user first attempts foraging search by inspecting all anchors, and all items within related anchors (given by \autoref{eq:forage}).
Upon not finding the item, a surprise penalty $T_c$ is incurred, and the user reverts to serial search under caution from the top of the menu.

To support foraging search, it is desirable for an adaptive system to create groups of related items, or eliminate groups where items have no associations.

\subsubsection{\textbf{Recall search}}

Recall (direct) search~\cite{bailly_14} relies on user's memory,
given by activations $B$, to directly glance at items without inspecting the entire menu.
For a target item $i$, if there are no activations $B_i$ above a threshold $\theta$ (we use 0.5),
the user reverts to serial search (\autoref{eq:serial}).
When $B(i_l) \geqslant \theta$ for a target $i_l$ at location $l$,
the user attempts recall search by inspecting the item at $l$ (\autoref{eq:read}).

If found at $l$, the user then performs a pointing task.
Here, the visual search and pointing task are performed sequentially as eye movement is faster than mouse movement \cite{bailly_14}. We use Fitts' law to estimate pointing time in menus:
\begin{equation}
    T_\text{pointing}(i_l) = a_p + b_p \cdot \log(1 + i_l)
\end{equation}
where $a_p=10.3$ and $b_p=4.8$ according to \cite{bailly_14}.

If not found at $l$, after incurring surprise penalty $T_c$, the user attempts \emph{local search}, by randomly inspecting $N_{local}$ items in the vicinity of location $l$.
\begin{equation}
    T_\text{local}(i_l) =  T_c + \delta \cdot  N_\text{local} + T_\text{trail}
\end{equation}
In non-ordered menus, $N_\text{local}$ is equal to 2 times the number of items in the Fovea.
In semantic  menus, $N_\text{local}$ is the number of items in the group.

In an adaptive menu, the target item $i$ might have been encountered at several locations.
Here, the user attempts recall search at all locations $l \in L$ where $B(i_l) \geqslant \theta$, until the target is found.
The total recall search time for target $i_a$ at adapted location $a$ is given by:
\begin{equation}
T_\text{recall}(i_a) = T_\text{pointing}(i_a) + \sum_{l\in L:B(i_l)\geqslant\theta}^{l \pm N/2} T_\text{read}({i_l}) + T_\text{local}(i_l)
\end{equation}

If recall search fails ($a \notin L$ or $B(t,a) < \theta$), the user eventually reverts to serial search under caution (\autoref{eq:serial-slow}).

To support recall, it is advantageous for an adaptive system to place frequently-encountered items at locations where they have been seen before.

\subsection{Neural Network for Rewards Estimation}
\label{sec:rnn-value-network}
\begin{figure}[b!]
    \centering
    \includegraphics[width=0.99\linewidth]{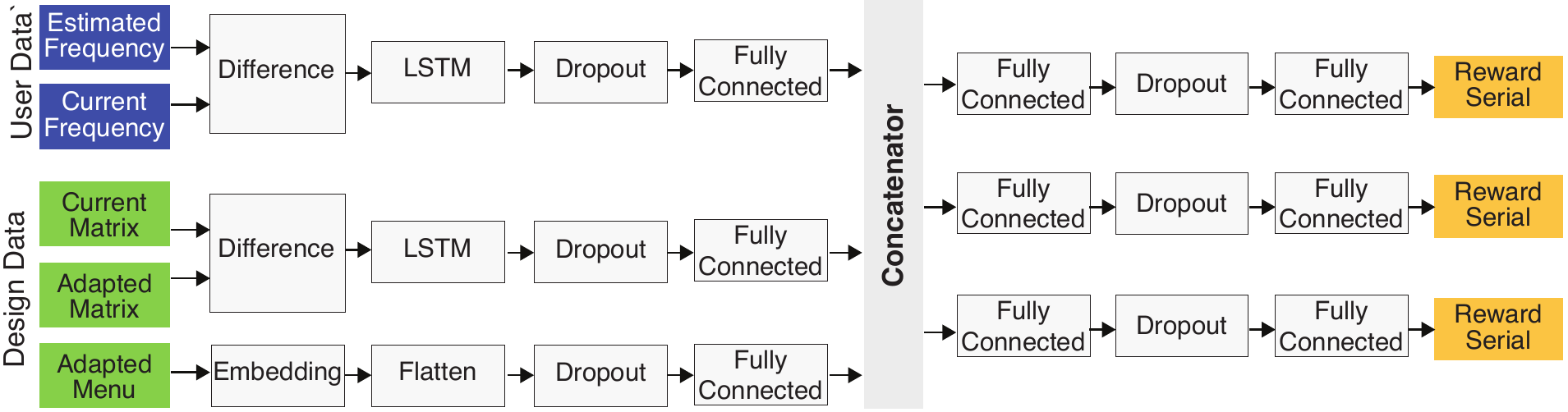}
    \caption{Our value network architecture takes menu design and user features as input to provide individual reward predictions.}
    \label{fig:value_network_menus}
     \Description{An architecture diagram for the neural network designed for the adaptive menu case. The diagram visualises the components that are described in the corresponding text.}
\end{figure}

The above search models enable us to predict selection times for varying user strategies.
By simulating consequences of adaptations during roll-outs,
we can estimate implications of design changes on user performance.
However, given the varying length of menu sizes,
running simulations with long horizons can be infeasible for online settings.
For example, for a menu with 15 items, and up to 8 separators,
there are over 500 feasible adaptations.
To address large problem sizes,
we instantiate our general network architecture (\autoref{fig:value_network_general})
for adaptive menus.
The key idea is to anticipate the rewards for a given menu adaptation
taking into account the previous menu design and the user behaviour.
The model inputs are:
(1)~design head: adapted menu design, association matrices of the current and adapted menu;
(2)~user head: previous and current clicks distribution.
The model outputs reward predictions for each of the three search models: serial, foraging, and recall.
\autoref{fig:value_network_menus} illustrates the model architecture.

Each input is treated as an independent model branch (head) that is eventually concatenated
and passed to three independent model branches (tails), one for each output reward.
Each item in the adapted menu is converted to a one-hot encoded vector, flattened, and then passed to a fully connected layer.
The association matrices are diffed and passed to a long short-term memory (LSTM) layer,
then passed to a fully connected layer.
Finally, the click history is passed to an LSTM layer,
which models sequential data and is designed to handle long-term dependencies,
and is then passed to a fully connected layer.

The concatenated inputs are passed to each network tail,
which comprises two stacked fully connected layers.
At the end of each tail, we use linear activation to predict each reward,
since they are not bounded.
For regularisation purposes, our architecture uses Dropout layers with drop rate $0.5$
before each of the fully connected layers.
This prevents overfitting the model to the training data, improving generalisability to unseen data.
The loss function for all model tails is the mean squared error (MSE),
which is computed as the average of the squared differences
between the predicted and the actual values, which penalises large errors.
We use the RMSProp optimiser, a popular stochastic gradient descent algorithm
with momentums.
We use learning rate $\eta=0.001$ and decay factor $\beta=0.9$ for the optimiser.
We train the model for 200 epochs at most, using early stopping (10 epochs patience)
to retain the best model weights, and monitor its performance on a validation set
comprising 20\% of training data.
After training, our model achieved remarkable performance:
$\text{MSE}_{serial} = 0.149$,
$\text{MSE}_{forage} = 0.408$,
$\text{MSE}_{recall} = 0.431$.

\section{Evaluation}

We validate our method, applied to adaptive menus, through technical and empirical evaluations.

\subsection{Technical Evaluation}
We conducted a technical evaluation with realistic and challenging scenarios,
where the adaptive system must adapt menus for simulated users.
The two main questions we seek to answer are:
\begin{enumerate}\compresslist
\item \emph{Can a model-based planning approach successfully and consistently improve predicted usability?}
\item \emph{Does our neural network based solution scale it up to address larger problem sizes?}
\end{enumerate}

\subsubsection{Tasks}\hspace*{\fill} \\
\emph{Menu Designs and User Interest:}
We considered 3 \emph{menu sizes} --- 5, 10, and 15 items --- to address varying cases, from short contextual menus to longer application menus.
In addition, up to 8 separators were allowed for grouping, resulting in menus with up to 23 items.
We picked common labels for menu items, where categories specified pairwise associations (e.g. animals, furniture, vegetables, clothing).
For each menu size, we created 4 starting menu designs by randomly assigning labels to item positions.
We used a Zipfian distribution to reasonably model menu usage \cite{Liu_frequency,cockburn_predictive_2007}. We sampled 8 different click histories by randomly assigning frequency to item labels.
This resulted in 3$\times$8$\times$4 = 96 configurations, each assigned to a different simulated user.

\emph{Reward Estimation:}
We compare two methods of estimating rewards: \emph{model-based simulations} and \emph{value neural network predictions}.
With model-based simulations, predictive models are used during roll-outs to estimate rewards for each state.
With value network predictions, our pre-trained network models are used to predict value estimates for each state.

\subsubsection{Implementation}
Our MCTS-based planning algorithm, and the predictive menu models, are implemented in Python 3.7.
The value neural network is implemented with TensorFlow 2.0.0.
We used a GNU/Linux server with an Intel Xeon Gold CPU @ 2.10\,GHz (64 bit processor) for simulations.
The execution of the study was automated such that trials were conducted sequentially, to avoid variations in computational resource usage.

During each trial, a combination of \{menu size $\times$ user history  $\times$ menu design $\times$ objective function $\times$ reward source\} was selected, and given as input to the system.
In a constrained setting, the MCTS algorithm was allowed 400 iterations, and a shallow roll-out horizon of 4 steps, to build the search tree and find suitable adaptations.

\begin{figure}
    \centering
    \includegraphics[width=0.95\linewidth]{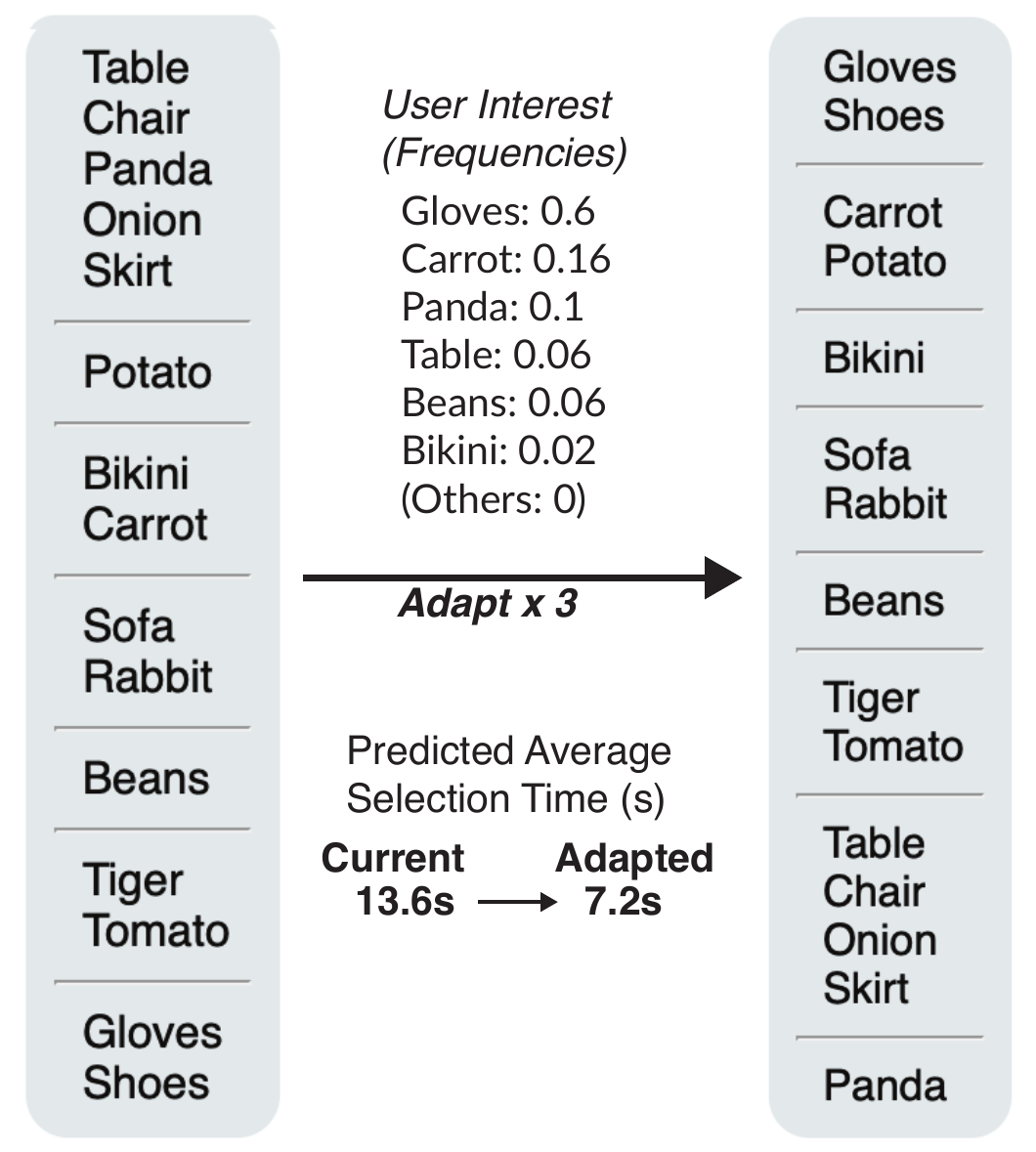}
    \caption{A sample result from the simulation study, for a 15-item menu design. In 3 steps, the menu was adapted to better suit the given user's interests (`Gloves' group moved to the top), and improved some grouping (`Carrot' with `Potato') as well. }
    \label{fig:result_menu_adapt}
     \Description{The figure shows an original menu on the left, and a final adapted menu on the right. An arrow connects the two menus, and the text above the arrows mentions the click frequencies on each item. The text below the arrow mentions that the menu is adapted 3 times, and also mentions that the original average selection time was 13.6 seconds while the final selection time is 7.2 seconds.}
\end{figure}

\subsubsection{Result: Success Rate}
With the above setup, we first evaluated whether our approach, and implementation, could successfully identify promising adaptations.
As dependent variable, we measured \emph{success rate} of finding beneficial adaptations.
We define a successful trial as one where the predicted selection time is improved by adaptation.
\autoref{fig:result_menu_adapt} shows an example result for a challenging case with a 15-item menu.

The overall success rate with model-based simulation was 92.7\%, indicating that in most cases an improvement was found.
Similarly, with the value network, success rate was 89.6\%.
These results support our approach towards planning menu adaptations that can improve user performance.

\subsubsection{Result: Scalability}

To assess the scalability of our solution, we compared computation time for model-based simulations vs. value network predictions.
In addition to the horizon of 4 steps used to evaluate success rate, we evaluated 3 other search depths -- 6, 8, and 10 -- depicting a range of planning horizons, from short sequences to longer sequences.

\begin{figure}
    \centering
    \includegraphics[width=.49\textwidth]{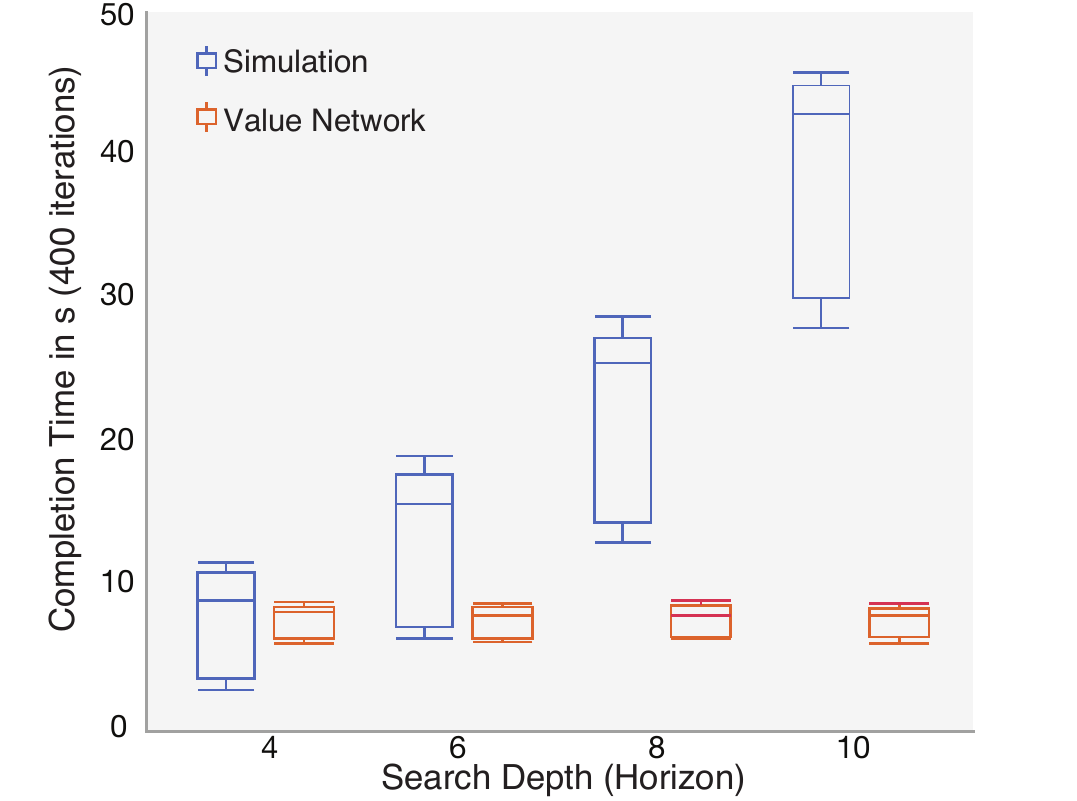}
    \caption{Computation time for planning adaptations via model-based simulations vs. value network for varying search depths. Our value network exhibited consistent performance for longer horizons.}
     \Description{A graph comparing computation time for model-based simulations vs. value network estimation. The four box plots for simulation results show a linear trend, increasing with search depth, while the four box plots for value network show that time remains constant with increasing depth.}
    \label{fig:results_vn}
\end{figure}

\autoref{fig:results_vn} illustrates computation time results for each depth level (for 400 MCTS iterations). We observe that for depths $\leq 4$, the value network does not provide much benefit.
However, as search depth increases, while the computation time with our value network remains constant (mean M = 7.77s, SD = 1.0), it drastically increases with simulations (from M = 7.9s, SD = 3.5s at depth 4 to M = 39.0s, SD = 7.4 at depth 10).

\subsection{Empirical Evaluation}

The primary goal of this evaluation is to test whether our planning approach (henceforth {\sc MCTS}) applied to linear menus improves performance in comparison with static menus ({\sc Static}),
and with the well-known frequency-based adaptive approach ({\sc Frequency}) as a baseline (e.g. as in \cite{lee2004quantitative}).
In {\sc MCTS}, the menu adapts after each block by planning adaptations; in {\sc Static}, the menu does not adapt over time; in {\sc Frequency}, the menu adapts based on the frequency of clicks on menu items.
To this end, we conducted a lab study where participants completed selection tasks in a \emph{within-subject design} with three conditions ({\sc Static, Frequency, MCTS}).

\subsubsection{Materials}
For the experiment, linear menus with 15 item labels were randomly generated.
Items labels were selected from common categories (e.g. animals, fruits, countries, etc.) to avoid prior biases.
For each participant, two menus were generated for each of the three conditions, resulting in six unique menus.
To avoid confusion, there were no overlaps in item labels or categories between the menus for a participant.

For each menu within a condition, a Zipfian distribution, known to accurately capture real-world command selections \cite{Liu_frequency,cockburn_predictive_2007}, with shape $s$ = 1.5 was used to control the frequency distribution of target items.
The same frequency distribution was used for all three conditions within a participant.
Unique frequency distributions were generated for every participant, to consider variance in user interests.
These frequency distributions were then used to generate sequences of target items, to be presented as stimulus during the trials.

\subsubsection{Participants}
18 participants (10 masculine, 8 feminine, 0 others), aged 18 to 38 (mean 27.2), with varying educational backgrounds, were opportunistically recruited.
All participants reported frequent desktop or mobile, and web usage.
Participation was compensated with a movie ticket voucher (approx. €12.00).

\subsubsection{Apparatus}
The experiment was conducted on a Macbook Pro, with a 15'' Retina display.
An Apple Magic Mouse with default tracking speed was used for selection tasks.
The study interface was implemented using HTML and Javascript, and was displayed in a browser window.
Timestamped cursor movements and clicks on menu items were recorded.

\subsubsection{Stimulus and Task}
The target item name was displayed at the top of the browser window.
Participants began a trial by clicking on a confirm button, upon which the menu was displayed directly below.
Errors were logged, and participants had to select the target item to finish the trial.
Upon clicking the target item, the menu was hidden, and a short break was provided.

\subsubsection{Procedure and design}
The experiment began with an introductory briefing and participant consent.
In a within-subject experimental design, each participant tested the three conditions ({\sc Static}, {\sc Frequency}, {\sc MCTS}) sequentially.
Condition order was counterbalanced between participants using a 3$\times$3 Latin square.

During each condition, the participant interacted with two different menus during 3 blocks.
Within a block, the two menus appeared in an alternating order, separated by short breaks (3 seconds).
For each menu, 20 selection tasks (trials) were completed.
We introduced this design to reflect the fact that
(1) users perform several sessions of work in the real life (one session == 1 block),
(2) users regularly switch between applications within a session, and thus use different menus \cite{raissi_retroactive_transfer}: we wanted to avoid undesirable effects due to repetitive selection within a single menu, and
(3) each menu has a different selection frequency, given by two Zipfian distributions.

Participants took mandatory breaks (1 minute) between two consecutive blocks, and longer breaks (5 minutes) between conditions where they answered open-ended interview questions.
In summary, the design is: 18 participants $\times$ 3 conditions $\times$ 3 blocks $\times$ 2 menus $\times$ 20 selections = 6480 trials.

\subsubsection{Quantitative Results}\hspace*{\fill} \\
\begin{figure*}
    \centering
    \includegraphics[width=0.9\textwidth]{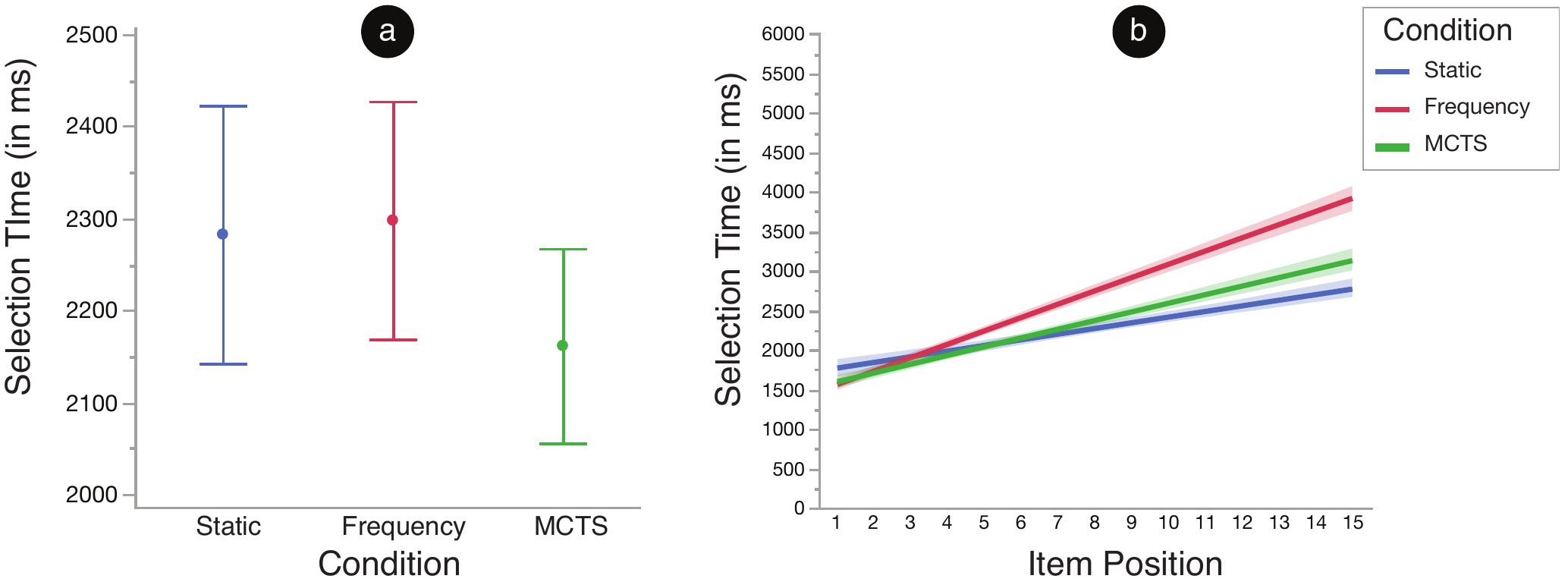}
    \caption{(a) The {\sc MCTS} condition was associated with significantly lower selection time as compared to the two baselines. Vertical bars indicate 95\% confidence intervals.
    (b) For items positioned lower in the menu, selection time in {\sc Frequency} increased more drastically than other conditions.}
     \Description{A two-part figure showing quantitative results from the user study. The left side shows a graph comparing the mean selection times for the 3 conditions; the right side shows linear fits of selection time vs. item position for the 3 conditions.}
    \label{fig:quant_results}
\end{figure*}
\emph{Average Selection Time:}
We created a mixed-effect model for repeated measures analysis of variance (one-way ANOVA) with selection time (in ms) as dependent variable, conditions ({\sc Static, Frequency, MCTS}) as fixed independent variable, and participant ID and menu design as random variables.

\emph{Condition} had a statistically significant effect on selection time $F(2,17)$ = 5.47, $p$ < 0.05, with grand means {\sc Static} = 2283 ms, {\sc Frequency} = 2298 ms, and {\sc MCTS} = 2162 ms (\autoref{fig:quant_results}a).
Post-hoc test using Tukey HSD revealed that {\sc MCTS} (2162 ms) was significantly faster (lower selection time) than both {\sc Frequency} (2298 ms) and {\sc Static} (2283 ms); the difference between {\sc Static} and {\sc Frequency} was not statistically significant.

\emph{Target Item Position:}
Given the menu selection scenario, items near the top of the menu are typically faster to select than items that are near the bottom.
However, this selection time depends not only on the cursor movement distance, but also the user's ability to search for items in the menu.
To get a better understanding of how the different conditions influenced performance, we further looked at how target item positions in the menu influence selection time.
\autoref{fig:quant_results}b illustrates the linear increase in selection time with target position for the three conditions.
It can be observed that while selection time for the top-most items (lower target positions) is quite similar for the three conditions, with {\sc MCTS} being the fastest, it increases more drastically for the {\sc Frequency} condition, as compared to {\sc Static} and {\sc MCTS}.
When we exclude the top-three target items, the difference in selection time between {\sc MCTS} (mean = 2454 ms) and {\sc Frequency} (mean = 2799 ms) is 344 ms (i.e.  {\sc Frequency} is about 15\% slower).

\subsubsection{Qualitative Results}
During the study, participants were not informed about adaptations (if any) in advance.
After each block, we asked them whether they noticed changes to menus during use, and their opinions about these changes (if any).
15 participants commented that they noticed changes in the {\sc Frequency} condition, but only 2 participants noticed \emph{how} these changes were occurring.
Participants commented that the reordering was confusing, and prevented them from remembering item locations: \emph{``I might skip down instead of checking the top, and then go back to the top.''} (P3).
In the {\sc Static} condition, participants could use their memory to directly access some items, but commented on the lack of proper grouping: \emph{``the items were not consistent in their grouping, and they were not intuitively grouped''} (P11).
In the {\sc MCTS} condition, participants noticed that the categorisation of items into groups improved upon adaptation, and helped them in searching for related items: \emph{``the items were organised under categories often -- that helped select items''} (P18).

\subsection{Summary}
Results from our simulation-based evaluation offer evidence for our approach and technical solutions.
First, MCTS-based planning consistently proposes adaptations that could improve predicted performance (\autoref{fig:quant_results}a).
Second, as search depth increases, our results indicate that value network is more efficient for estimating reward predictions. Further, results from our user study highlight benefits over a static and an adaptive baseline.
Through model-based planning, we can adapt menus that improve overall performance, as given by reduced selection time.
A common pitfall of the frequency-based approach is that, although it can improve performance for commonly-used items or commands, it prevents recall and makes selection of other items increasingly difficult.
In contrast, we observed (\autoref{fig:quant_results}b) that adaptations made through our approach could provide these benefits while avoiding costly changes that require relearning and cause annoyance.

\section{Conclusion}

\begin{figure}[b!]
    \centering
    \includegraphics[width=0.46\textwidth]{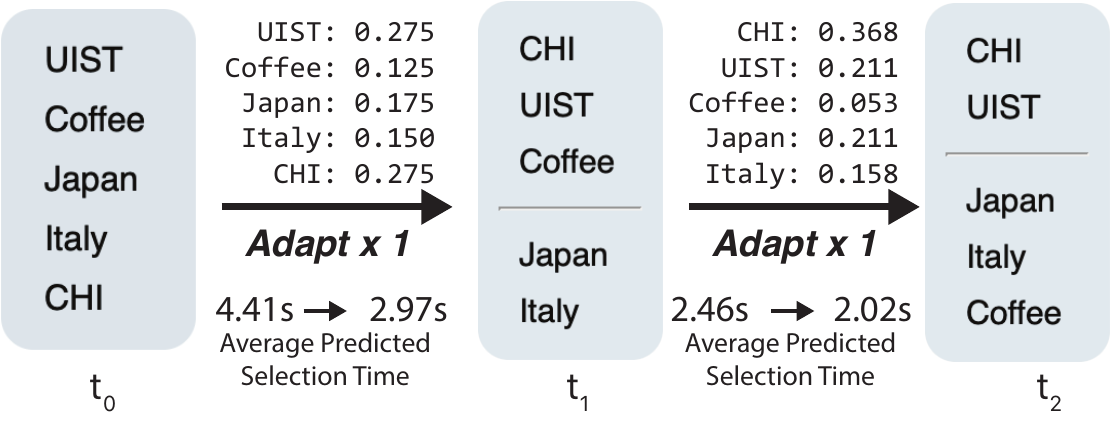}
    \caption{Sequential adaptation of a 5-item menu. The system avoids greedily moving `Coffee' (lowest frequency) at the first step. In light of new observations, this change is justified by the reduced user interest.}
     \Description{A three-part figure showing two adaptations of a 5-item menu. Each part shows the menu structure, and the parts are connected with arrows that mention the click frequency and change in predicted selection time.}
    \label{fig:result_5items}
\end{figure}

We have presented a model-based reinforcement learning method suitable for adaptive interactions.
While recent successes of reinforcement learning have created renewed enthusiasm toward this approach in HCI, applications to adaptive user interfaces have remained scarce.
To apply this class of machine learning methods for selecting adaptations,
we have proposed the use of predictive models in HCI for value estimation during planning.
We have presented solutions to several consequent technical challenges, most notably:
\begin{itemize}
\item How to model the decision problem in adaptive interfaces for model-based RL;
\item How to estimate MCTS roll-outs using HCI models;
\item How to design deep neural networks to boost planning.
\end{itemize}

To study and demonstrate the viability of the approach, we have applied it to the challenging case of adaptive menus by extending predictive models.
Our simulated and empirical evaluations suggest significant and practically valuable improvements to usability.
Importantly, the adaptive system does not require explicit user input,
but is still able to perform conservatively without disadvantageous or annoying changes (\autoref{fig:result_5items}).
Our empirical evaluation reveals that this approach can work even when starting from poor designs that would be hard to recover with approaches that do not consider planning.
Future adaptive applications can benefit from our general approach by exploiting and extending predictive models of interaction.

\subsection*{Limitations and Future Work}

We see several exciting topics for future research on model-based RL and its applications in HCI.
First, one limitation to the applicability of the approach is the requirement for accurate models of short- and long-term consequences of adaptations.
So far we have assumed that such models are expressed as step-by-step computer simulations or mathematical models.
However, there is no reason why data-driven models (e.g. \cite{10.1145/3313831.3376870})  -- trained on larger datasets of user data -- could not be used for this purpose, significantly expanding the scope of possible applications.
Second, algorithm engineering is needed to deploy this approach to larger applications.
In particular, presently, with our computing resources,
problem sizes of up to 20 items are still within reach in the case of menu systems.
To improve performance beyond that,
techniques for GPU computation and more efficient training will need attention.
Finally, while our work successfully used a value network,
further improvements can be expected by implementing a policy network \cite{silver2016mastering}.

To conclude, we hope our work can be broadly appealing, and invite contributions from both the HCI and the machine learning community.
At the core of model-based RL is an understanding of how users behave
and what makes a good design in given conditions.
We believe that future applications can benefit from this approach to improve interactions.

\urlstyle{tt}
\definecolor{linkColor}{RGB}{20,100,245}

\section{Open Science}
We support adoption and further research efforts by providing an open code repository, with examples and instructions, on our project page: \textcolor{linkColor}{\url{https://userinterfaces.aalto.fi/adaptive}}.

\urlstyle{rm}

\begin{acks}
We thank all study participants for their time, and colleagues and reviewers for their helpful comments.
This work was funded by the Department of Communications and Networking (Comnet), the Finnish Center for Artificial Intelligence (FCAI), Academy of Finland projects `Human Automata' and `BAD', Agence Nationale de la Recherche (grant number ANR-16-CE33-0023), and HumaneAI Net (H2020 ICT 48 Network of Centers of Excellence).
\end{acks}

\bibliographystyle{ACM-Reference-Format}
\bibliography{references}

% %%
% %% If your work has an appendix, this is the place to put it.
% \appendix

% \section{Appendix 1}

\end{document}